\documentclass{article}
\usepackage[dvips]{graphicx}
\usepackage{spconf,amssymb,amsmath,epsfig,color}
\newcommand{\beq}{\begin{equation}}
\newcommand{\eeq}{\end{equation}}
\newcommand{\beqn}{\begin{eqnarray}}
\newcommand{\eeqn}{\end{eqnarray}}
\DeclareMathOperator*{\argmin}{arg\,min}
\def\bmath#1{\mbox{\boldmath$#1$}}

\long\def\symbolfootnote[#1]#2{\begingroup%
\def\thefootnote{\fnsymbol{footnote}}\footnote[#1]{#2}\endgroup}

\title{Distributed Model Construction in Radio Interferometric Calibration}
\name{Sarod Yatawatta}
\address{ASTRON, The Netherlands Institute for Radio Astronomy,\\ The Netherlands.\\ Email: yatawatta@astron.nl}
\begin{document}
\ninept
\maketitle
\begin{abstract}
Calibration of a typical radio interferometric array yields thousands of parameters as solutions.  These solutions contain valuable information about the systematic errors in the data (ionosphere and beam shape). This information could be reused in calibration to improve the accuracy and also can be fed into imaging to improve the fidelity. We propose a distributed optimization strategy to construct models for the systematic errors in the data using the calibration solutions. We formulate this as an elastic net regularized distributed optimization problem which we solve using the alternating direction method of multipliers (ADMM) algorithm. We give simulation results to show the feasibility of the proposed distributed model construction scheme.
\end{abstract}
\section{Introduction}
Radio interferometric observations are almost always affected by systematic errors. During calibration, these systematic errors are estimated along many directions in the sky using compact celestial sources as guide beacons. In addition, the corrupting signals are also subtracted from the data to reveal weaker signals of interest. Large volumes of data need to be calibrated to deliver the science goals of modern radio astronomy. As a secondary outcome of calibration, hundreds of thousands of parameters are obtained as  calibration solutions and are stored as metadata.

The main sources of systematic errors in radio interferometric data are the ionosphere and the receiver beam shape. The effect of ionosphere is mostly represented as a phase (or total electron content) screen in radio astronomy \cite{Cotton,Intema} as well as in other applications such as space weather \cite{GNSS}. The calibration solutions along the directions of compact sources are used to build phase screens \cite{Cotton,Intema,Mevius} and this is further improved  to operate in real time \cite{RTS}. Faraday rotation of incoming radiation is an additional complication caused by the ionosphere (that can be seen with dual polarized antennae) and  external information such as GNSS (global navigation satellite system) satellites are used to model this \cite{Faraday,arora,arora2}. It is noteworthy that phase screen models have limited accuracy for science goals that demand high dynamic ranges \cite{Martin2016}.

The receiver beam shape has traditionally been estimated using holographic techniques \cite{Scott,Bennet,Harp}. Recent surge in the use of unmanned areal vehicles (drones) have enabled their use in beam shape estimation as well \cite{Chang15,Jacobs}. It is also possible to use calibration solutions to obtain beam models \cite{BEAM}. Once accurate models for the ionosphere and the beam shape have been obtained, image fidelity can be improved by incorporating such models into the imaging process \cite{WBA,Aproj}.

In this paper, we propose a method to construct a unified model for the ionosphere and the beam shape. The novelty (relation to prior work) is as follows: (i) We extend the scalar models (single polarization) \cite{Cotton,Intema,RTS,BEAM} to handle data taken with dual polarized receivers. Our model incorporates the refraction and the Faraday rotation due to the ionosphere into one. (ii) We create a unified model for both the ionosphere and the beam shape and can be directly used by imaging algorithms \cite{WBA,Aproj}. We enforce elastic net regularization \cite{ElasticNet} during model creation. The power constraint comes naturally because the received signals have finite power \cite{BEAM}. The sparseness is more subtle, but arises because the relative difference in the ionosphere seen by receivers close together on Earth is small \cite{Lonsdale,Tol,Wijn17}. Therefore, sparseness minimizes overfitting, for example when there is no differential Faraday rotation between stations. (iii) We propose a distributed optimization scheme using the alternating direction method of multipliers (ADMM) \cite{boyd2011}. This scheme matches nicely with the distributed calibration schemes in use \cite{DCAL,Brossard2016}, and also makes our algorithm computationally efficient. 
 
The rest of the paper is organized as follows: We give a brief introduction to radio interferometric calibration and models for systematic errors in section \ref{sec:model}. We propose a distributed model construction method based on consensus optimization in section \ref{sec:dmod}. We show the feasibility of the proposed method in  section \ref{sec:simul} using simulated data before drawing our conclusions in section \ref{sec:conclusions}.
Notation: Matrices and vectors are denoted by bold upper and lower case letters as ${\bf J}$ and ${\bf v}$, respectively. The transpose and the  Hermitian transpose are given by $(\cdot)^T$ and $(\cdot)^H$, respectively. The matrix  Frobenius norm is given by $\|\cdot \|$ and the $l_1$ norm by $\|\cdot \|_1$. The set of real and complex numbers are denoted by  ${\mathbb R}$ and ${\mathbb C}$, respectively. The identity matrix (size $N\times N$) is given by ${\bf I}_N$.

\section{Radio Interferometric Data Model}\label{sec:model}
We consider an array with $N$ stations and the observed data ${\bf V}_{pqf} \in \mathbb{C}^{2\times 2}$ at the baseline formed by stations $p$ and $q$  at frequency $f$ is given by \cite{HBS} 
\beq \label{ME}
{\bf V}_{pqf}=\sum_{k=1}^{K} {\bf J}_{pkf} {\bf C}_{pqkf} {\bf J}_{qkf}^H + {\bf N}.
\eeq
The data consists of the signals from $K$ sources in the sky ${\bf C}_{pqkf}\in \mathbb{C}^{2\times 2}$ $p,q\in[1,N]$, $k\in[1,K]$ corrupted by the systematic errors ${\bf J}_{pkf},{\bf J}_{qkf} \in \mathbb{C}^{2\times 2}$. The systematic errors represent the cumulative effect of the ionosphere and the beam shape and also the receiver electronics. We consider the data (taken at $F$ distinct frequencies) to be stored across a network of computers. Using distributed calibration \cite{DCAL}, we estimate the systematic errors ${\bf J}_{pkf}$ for all $p,k$ and $f$ and we also store the solutions across the network of computers. 

Provided that the $K$ directions are spread across the full field of view, our objective is to create a model for the systematic errors across the field of view of each station. We use $G$ basis functions to cover the field of view and these basis functions represent the variation of the systematic errors both spatially as well as with frequency. Given that there are $F\times K$ solutions per each station, we assume $G\ll F\times K$. Let the $k$-th direction have spatial coordinates $(\alpha_k,\beta_k)$, and at frequency $f$, let the $i$-th basis function be $\phi_i(\alpha_k,\beta_k,f)$, $i\in[1,G]$. 

Based on the model ${\bf X}$ ($\in \mathbb{C}^{2N\times 2G}$) and the basis functions evaluated along the $k$-th direction at frequency $f$, $\bmath{\Phi}_{\alpha_k\beta_k f}$ ($\in \mathbb{C}^{2G\times 2}$), we can represent systematic errors for all $N$ stations along the $k$-th direction ${\bf J}_{kf}$ ($\in \mathbb{C}^{2N\times 2}$) as
\beq \label{mod}
 {\bf J}_{kf}={\bf X} \bmath{\Phi}_{\alpha_k\beta_k f}
\eeq
where
\beqn \label{basis} {\bf X}\buildrel\triangle\over = \left[
\begin{array}{cccc}
{\bf X}_{11} & {\bf X}_{12} & \ldots & {\bf X}_{1G}\\
{\bf X}_{21} & {\bf X}_{22} & \ldots & {\bf X}_{2G}\\
\ldots & \ldots & \ldots & \ldots\\
{\bf X}_{N1} & {\bf X}_{N2} & \ldots & {\bf X}_{NG}
\end{array} \right],\\\nonumber
\bmath{\Phi}_{\alpha_k\beta_k f} \buildrel\triangle\over=\left[
\begin{array}{c}
\phi_1(\alpha_k,\beta_k,f)\\
\phi_2(\alpha_k,\beta_k,f)\\
\ldots\\
\phi_G(\alpha_k,\beta_k,f)
\end{array} \right] \otimes {\bf I}_{2},\ 
{\bf J}_{kf} \buildrel\triangle\over=\left[
\begin{array}{c}
{\bf J}_{1kf}\\
{\bf J}_{2kf}\\
\ldots\\
{\bf J}_{Nkf}
\end{array} \right].
\eeqn
Each ${\bf X}_{pi}$ ($\in \mathbb{C}^{2\times 2}$) $p\in[1,N]$ $i\in[1,G]$ in ${\bf X}$ represents the contribution of the $i$-th  basis function towards the model of the systematic errors of the $p$-th station. Note that each ${\bf X}_{pi}$ is independent of spatial coordinates or frequency.

It seems straightforward to estimate ${\bf X}$ by augmenting many calibration solutions ${\bf J}_{kf}$ and inverting (\ref{mod}). However, the solutions attainable for (\ref{ME}) are ${\bf J}_{kf}{\bf U}_{kf}$ where ${\bf U}_{kf}$ ($\in\mathbb{C}^{2\times 2}$) is an unknown unitary matrix ${\bf U}_{kf}{\bf U}^H_{kf} ={\bf I}$. The reason for this unitary ambiguity is that ${\bf C}_{pqkf}$ in (\ref{ME}) is diagonal for most celestial sources. Notably, the unitary ambiguity will be different for each direction $k$ and for each frequency $f$. Therefore, we cannot use (\ref{mod}) directly to find ${\bf X}$.

\section{Distributed model construction}\label{sec:dmod}
In order to overcome the inherent unitary ambiguity, we reformulate our problem as follows. Taking the product
\beq \label{prod}
 {\bf J}_{pkf} {\bf C}_{pqkf} {\bf J}_{qkf}^H = {\bf A}_p {\bf X} \bmath{\Phi}_{\alpha_k\beta_k f} \widetilde{\bf C}_{pqkf} \left({\bf A}_q {\bf X} \bmath{\Phi}_{\alpha_k\beta_k f} \right)^H 
\eeq
using calibration solutions ${\bf J}_{pkf}$ and ${\bf J}_{qkf}$, we see that the unitary ambiguity cancels out because it is the same for both ${\bf J}_{pkf}$ and ${\bf J}_{qkf}$ and because ${\bf C}_{pqkf}$ is diagonal. In (\ref{prod}), ${\bf A}_p$ ($\in \mathbb{R}^{2\times 2N}$) is a matrix of zeros except at the $p$-th $2\times 2$ block it is ${\bf I}_2$,
\beq
{\bf A}_p\buildrel\triangle\over = [{\bf 0}\ {\bf 0}\ldots {\bf I}_2\ldots {\bf 0}]
\eeq
(and ${\bf A}_q$ likewise). The model for the $p$-th station is given by ${\bf A}_p {\bf X}$. The (updated) sky contribution $\widetilde{\bf C}_{pqkf}$  used in constructing the model need not be equal to ${\bf C}_{pqkf}$ which is used in calibration. Both $\widetilde{\bf C}_{pqkf}$ and ${\bf C}_{pqkf}$ are almost always diagonal matrices (because the sky signal is intrinsically unpolarized). We define a cost function as
\beqn \label{fcost}
\lefteqn{h({\bf X}) } &&\buildrel \triangle \over=  \sum_{pqkf} \|  {\bf J}_{pkf} {\bf C}_{pqkf} {\bf J}_{qkf}^H \\\nonumber
& & \mbox{} - {\bf A}_p {\bf X} \bmath{\Phi}_{\alpha_k\beta_k f} \widetilde{\bf C}_{pqkf} \left({\bf A}_q {\bf X} \bmath{\Phi}_{\alpha_k\beta_k f} \right)^H  \|^2
\eeqn
where the inclusion of $\widetilde{\bf C}_{pqkf}$ and ${\bf C}_{pqkf}$ in the cost function acts as a weighting, i.e., giving larger weights to solutions along the directions with large intensities, and therefore with higher signal to noise ratios. Also different $k$-s will have different fluxes, not all normalized in the input model, so this also acts as a normalization across all directions in the sky.

Let the data be partitioned into different frequency subsets and let the $j$-th partition contain frequencies given by the set $\mathcal{F}_j$. Each $\mathcal{F}_j$ is assumed to represent the data stored in one compute node. We separate the summation in (\ref{fcost}) as
\beq \label{fcostsum}
h({\bf X}) = \sum_j h_j({\bf X})
\eeq
where 
\beqn
\lefteqn{h_j({\bf X})}&&\mbox{}=\sum_{f\in \mathcal{F}_j}\sum_{pqk} \|  {\bf J}_{pkf} {\bf C}_{pqkf} {\bf J}_{qkf}^H \\\nonumber
& & \mbox{}- {\bf A}_p {\bf X} \bmath{\Phi}_{\alpha_k\beta_k f} \widetilde{\bf C}_{pqkf} \left({\bf A}_q {\bf X} \bmath{\Phi}_{\alpha_k\beta_k f} \right)^H  \|^2
\eeqn
correspond to the cost function local to the $j$-th compute node ($\sum_j$ implies summing over all compute nodes' cost functions).

The model is constructed by minimizing (\ref{fcost}) or (\ref{fcostsum}) with elastic net regularization \cite{ElasticNet}
\beq \label{origx}
{\bf X}=\underset{{\bf X}}{\argmin}\ \sum_j h_{j}({\bf X}) + \lambda\| {\bf X}\|^2 + \mu\| {\bf X}\|_1
\eeq
where $\lambda,\mu \in \mathbb{R}^{+}$.
Solving (\ref{origx}) directly is not tractable and noting that each $h_{j}({\bf X})$ is calculated on different compute nodes, we redefine (\ref{origx}) as a consensus problem \cite{boyd2011}
\beqn \label{consensus}
{\bf X}_1,{\bf X}_2,\ldots,{\bf Z}=\underset{{\bf X}_1,{\bf X}_2,\ldots,{\bf Z}}{\argmin}\ \sum_j h_{j}({\bf X}_j) + \lambda \| {\bf Z}\|^2 + \mu \|{\bf Z}\|_1\\\nonumber \mathrm{subject\ to\ } {\bf X}_j={\bf Z}\  \forall\ j,\ \ \mathrm{and}\  {\bf X}_1,{\bf X}_2,\ldots,{\bf Z} \in \mathbb{C}^{2N\times 2G}. 
\eeqn
The augmented Lagrangian for solving (\ref{consensus}) using ADMM is
\beqn \label{aug}
\lefteqn{L({\bf X}_1,{\bf X}_2,\ldots,{\bf Z},{\bf Y}_1,{\bf Y}_2,\ldots)}\\\nonumber
&&=\sum_j h_j({\bf X}_j) + \| {\bf Y}_j^H ({\bf X}_j-{\bf Z})\| + \frac{\rho}{2} \|{\bf X}_j-{\bf Z}\|^2\\\nonumber
&&+\lambda \|{\bf Z}\|^2+\mu \|{\bf Z}\|_1
\eeqn
where $\rho$ $\in \mathbb{R}^{+}$ is the penalty parameter and ${\bf Y}_j \in \mathbb{C}^{2N\times 2G}$ is the Lagrange multiplier local to compute node $j$.

The ADMM iterations for solving (\ref{consensus}) are (using $n=1,2,\ldots$ superscript for iteration number)
\beq \label{step1}
{\bf X}_j^{n+1} =\underset{{\bf X}_j}{\argmin}\  L({\bf X}_j,{\bf Z}^{n},{\bf Y}_j^n)
\eeq
\beq \label{step2}
{\bf Z}^{n+1} =\underset{{\bf Z}}{\argmin}\  L({\bf X}_1^{n+1},{\bf X}_2^{n+1},\ldots,{\bf Z},{\bf Y}_1^{n},{\bf Y}_2^{n},\ldots)
\eeq
\beq \label{step3}
{\bf Y}_j^{n+1}={\bf Y}_j^n + \rho({\bf X}_j^{n+1}-{\bf Z}^{n+1}).
\eeq
Steps (\ref{step1}) and (\ref{step3}) are performed in a distributed manner at each compute node and the intermediate step (\ref{step2}) is performed at the fusion center. There is no closed form solution for  (\ref{step1}) and we use the Riemannian trust region method \cite{RTR,manopt} to find a solution. The required gradient and Hessian operators are given in the appendix. 

A solution for (\ref{step2}) is obtained in closed form as
\beq
{\bf Z} = \bmath{\Psi}_{\frac{2 \mu}{2\lambda+\sum_{j^\prime} \rho}}\left(\frac{1}{2\lambda+\sum_{j^\prime} \rho} \sum_j\left({\bf Y}_j+\rho{\bf X}_j\right)\right)
\eeq
where $\bmath{\Psi}(\cdot)$ is the (matrix) soft threshold operator, its scalar version being $\Psi_{\lambda}(x)=\mathrm{sign}(x) \mathrm{max}(|x|-\lambda,0)$.

\section{Simulations}\label{sec:simul}
\begin{figure}[htbp]
\begin{minipage}[b]{0.98\linewidth}
\centering
\centerline{\epsfig{figure=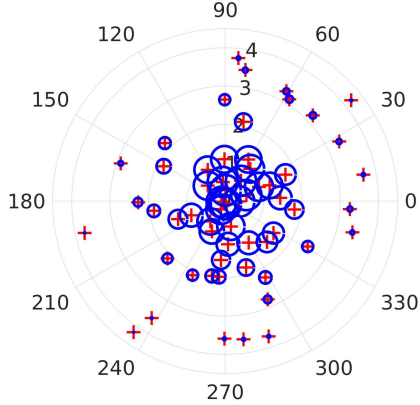,width=7.5cm}}
\end{minipage}
\caption{Sky model spread across a field of view of $10$ degrees in diameter. The blue circles are scaled according to the intensity of each source.} \label{fig_sky}
\end{figure}

We simulate an array with $N=16$ stations, collecting data at $F=10$ frequencies in the range $[60,180]$ MHz. Note that $F$ can be many hundreds in real observations. The sky consists of $K=60$ sources, spread across a field of view of $10$ degrees in diameter. The beam shapes of each $N$ stations are randomly generated Gaussians, with random pointing centers and footprints. The width of the beams are varied according to $1/f^2$. The source intensities are attenuated according to the average beam shape. As seen in Fig. \ref{fig_sky}, the sources at the center have higher intensities than at the edge of the field of view.
To simulate the effect of the ionosphere, each beam shape is multiplied by a complex number $\exp\jmath (a_1 \alpha/f + a_2 \beta/f)$ where $(\alpha,\beta)$ are spatial coordinates and $a_1,a_2$ are drawn from $\mathcal{U}(-5,5)$. Finally, each scalar beam shape is multiplied by a randomly generated rotation matrix $\in\mathbb{C}^{2\times 2}$ to simulate the effect of Faraday rotation (rotation angle scales as $1/f^2$).
Using this compound model, we calculate the calibration solutions ${\bf J}_{kf}$ and multiply them with a random unitary matrix $\in\mathbb{C}^{2\times 2}$. We also add noise (a random matrix $\in \mathbb{C}^{2N\times 2}$) to ${\bf J}_{kf}$ with a norm that is 5\% of $\|{\bf J}_{kf}\|$.

\begin{figure}[htbp]
\begin{minipage}[b]{0.98\linewidth}
\centering
\centerline{\epsfig{figure=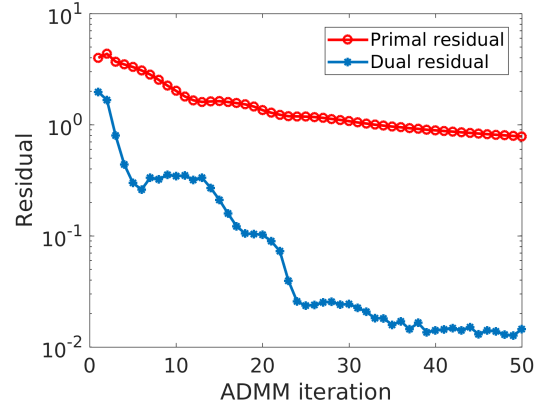,width=7.5cm}}
\end{minipage}
\caption{Variation of primal and dual residuals with ADMM iteration.} \label{fig_residuals}
\end{figure}

The basis functions $\bmath{\Phi}_{\alpha_k\beta_k f}$ in (\ref{basis}) are constructed by using $16$ spherical harmonics (for spatial dependence) multiplied with $5$ Bernstein bases (for frequency dependence). Therefore, $G=16\times 5=80\ll F\times K=10\times 60=600$. We use $50$ ADMM iterations, with penalty $\rho=10$ and regularization parameters $\lambda=40$ and $\mu=10$. We show the primal ($\|{\bf X}_j^n-{\bf Z}^{n}\|$) and  dual ($\|{\bf Z}^{n}-{\bf Z}^{n-1}\|$) residuals in Fig. \ref{fig_residuals}. We see that the primal residual is much higher, mainly because of the frequency dependence (Bernstein bases) not being representative enough.

\begin{figure}[htb]
\begin{minipage}[b]{0.98\linewidth}
\centering
\centerline{\epsfig{figure=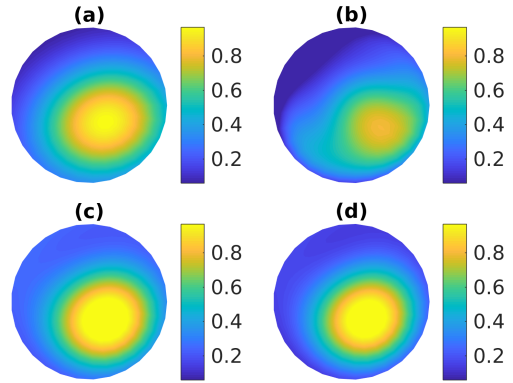,width=7.5cm}}
\end{minipage}
\caption{XX systematic error real part across the full field of view: (a) ground truth (b) linear estimate (c) consensus without regularization (d) consensus with regularization.\label{fig_XX_real}}
\end{figure}

\begin{figure}[htb]
\begin{minipage}[b]{0.98\linewidth}
\centering
\centerline{\epsfig{figure=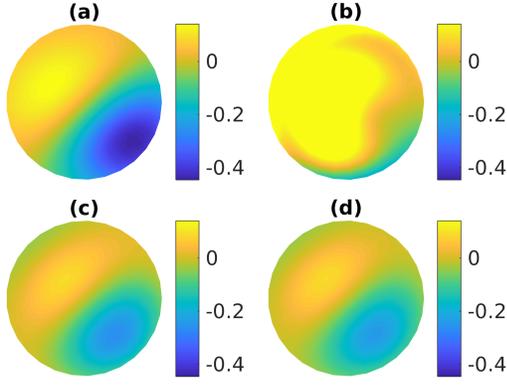,width=7.5cm}}
\end{minipage}
\caption{XX systematic error imaginary part across the full field of view: (a) ground truth (b) linear estimate (c) consensus without regularization (d) consensus with regularization.\label{fig_XX_imag}}
\end{figure}

We compare the solution obtained by (\ref{consensus}) with the linear estimate obtained by solving (\ref{mod}). Note that the linear estimate is always inferior to the solution obtained by (\ref{consensus}) because of the unitary ambiguities. We show the systematic error models constructed for one station in Figs. \ref{fig_XX_real} and \ref{fig_XX_imag}, showing the real and imaginary parts  of the systematic errors for one correlation (XX). As expected, the consensus optimization based solution (with and without regularization) gives better results than the linear estimate.  

\begin{figure}[htb]
\begin{minipage}[b]{0.98\linewidth}
\centering
\centerline{\epsfig{figure=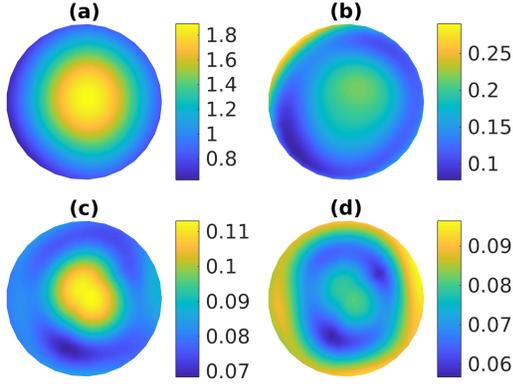,width=7.5cm}}
\end{minipage}
\caption{Average model construction error across the full field of view: (a) norm of ${\bf J}_{kf}$ (b) linear estimate (c) consensus without regularization (d) consensus with regularization.\label{fig_nmse}}
\end{figure}

In order to study the effect of elastic net regularization, we calculate the ground truth value of systematic errors ${\bf J}_{kf}$ and the estimated systematic errors based on the constructed model $\widehat{{\bf J}}_{kf}$ and find the difference (subject to a unitary ambiguity ${\bf U}$ \cite{interpolation})  as $\|{\bf J}_{kf} - \widehat{{\bf J}}_{kf} {\bf U} \|$. We call this {\em model construction error}. We evaluate the model construction error on a spatial grid of  $30\times 30$ directions (covering the full field of view) and average this over all $N$ stations. The model construction error surface over the full field of view at $f=100$ MHz is shown in Fig. \ref{fig_nmse}. For comparison, we have also shown the average $\| {\bf J}_{kf} \|$ in Fig. \ref{fig_nmse} (a).  We see the improvement due to elastic net regularization by comparing Fig. \ref{fig_nmse} (c) (no regularization) with Fig. \ref{fig_nmse} (d). The number of nonzero elements in the model ${\bf X}$ in the case of Fig. \ref{fig_nmse} (c)  is $2N\times 2G=5120$ while with elastic net regularization, this value becomes $4131$.

\begin{figure}[htb]
\begin{minipage}[b]{0.98\linewidth}
\centering
\centerline{\epsfig{figure=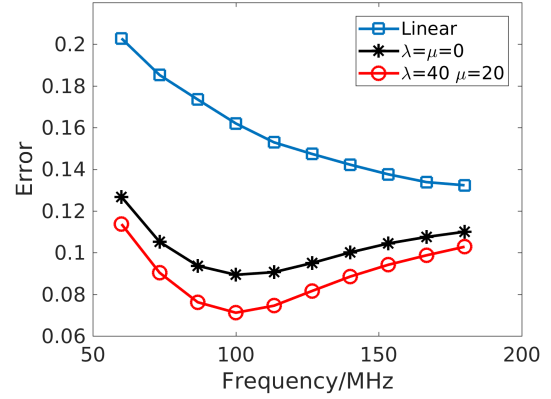,width=7.5cm}}
\end{minipage}
\caption{Average model construction error variation with frequency.\label{fig_nmse_all}}
\end{figure}

The variation of the average model construction error (over the full field of view) with frequency is shown in Fig. \ref{fig_nmse_all}. We can clearly see the improvement due to consensus optimization and elastic net regularization in Fig. \ref{fig_nmse_all}.

\section{Conclusions}\label{sec:conclusions}
We have formulated the construction of models for systematic errors in radio interferometric data as a distributed optimization problem. We solve this problem with the use of the ADMM algorithm and with elastic net regularization. Simulations show the feasibility of the proposed algorithm as well as the improvement gained by the elastic net regularization. Future work will focus on the application of this method to real observations and developing distributed software that increases computational speed.

\begin{center}
{\bf APPENDIX: GRADIENT AND HESSIAN\vspace{-.5em}\vspace{0pt}}
\end{center}

The form of the original cost function (\ref{fcost}) is structurally similar to the one considered in \cite{DCAL}. Therefore, by simple substitutions, it is possible to derive the gradient and the Hessian. We get the gradient of the augmented Lagrangian as 
\beq
\mathrm{grad}(L,{\bf X}_j) = \mathrm{grad}(h_j({\bf X}_j),{\bf X}_j) + \frac{1}{2}{\bf Y}_j + \frac{\rho}{2}({\bf X}_j-{\bf Z})
\eeq
where
\beqn
\lefteqn{\mathrm{grad}(h_j({\bf X}),{\bf X})}\\\nonumber
&&=-\sum_{f\in \mathcal{F}_j}\sum_{pqk} \left( {\bf A}_p^T \left( {\bf J}_{pkf} {\bf C}_{pqkf} {\bf J}_{qkf}^H \right.\right. \\\nonumber 
&&- \left.\left. {\bf A}_p {\bf X} \bmath{\Phi}_{\alpha\beta f} \widetilde{\bf C}_{pqkf} \bmath{\Phi}_{\alpha\beta f}^H {\bf X}^H {\bf A}_q^T\right) {\bf A}_q{\bf X} \bmath{\Phi}_{\alpha\beta f} \widetilde{\bf C}_{pqkf}^H \bmath{\Phi}_{\alpha\beta f}^H \right.\\\nonumber
&&+{\bf A}_q^T \left({\bf J}_{pkf} {\bf C}_{pqkf} {\bf J}_{qkf}^H \right.\\\nonumber
&&-\left.\left.{\bf A}_p {\bf X} \bmath{\Phi}_{\alpha\beta f} \widetilde{\bf C}_{pqkf} \bmath{\Phi}_{\alpha\beta f}^H {\bf X}^H {\bf A}_q^T\right)^H {\bf A}_p{\bf X} \bmath{\Phi}_{\alpha\beta f} \widetilde{\bf C}_{pqkf} \bmath{\Phi}_{\alpha\beta f}^H \right).
\eeqn
Similarly, the Hessian becomes
\beq
\mathrm{Hess}(L,{\bf X}_j,\bmath{\eta})=\mathrm{Hess}(h_j({\bf X}_j),{\bf X}_j,\bmath{\eta}) + \frac{\rho}{2}\bmath{\eta}
\eeq
where
\beqn
\lefteqn{\mathrm{Hess}(h_j({\bf X}),{\bf X},\bmath{\eta})}\\\nonumber
&=&\sum_{f\in \mathcal{F}_j}\sum_{pqk}\left( {\bf {A}}_p^T \left( ({\bf J}_{pkf} {\bf C}_{pqkf} {\bf J}_{qkf}^H \right.\right.\\\nonumber
&&-\left.\left.{\bf {A}}_p{\bf {X}}\bmath{\Phi}_{\alpha\beta f} \widetilde{\bf C}_{pqkf} \bmath{\Phi}_{\alpha\beta f}^H {\bf {X}}^H{\bf {A}}_q^T) {\bf {A}}_q {\bmath \eta}\right.\right.\\\nonumber
&& \left.\left.- {\bf {A}}_p({\bf {X}} \bmath{\Phi}_{\alpha\beta f} \widetilde{\bf C}_{pqkf} \bmath{\Phi}_{\alpha\beta f}^H {\bmath \eta}^H \right.\right.\\\nonumber
&&\left.\left.+ {\bmath \eta}\bmath{\Phi}_{\alpha\beta f} \widetilde{\bf C}_{pqkf} \bmath{\Phi}_{\alpha\beta f}^H {\bf {X}}^H) {\bf {A}}_q^T{\bf {A}}_q{\bf {X}}\right) \bmath{\Phi}_{\alpha\beta f} \widetilde{\bf C}_{pqkf}^H \bmath{\Phi}_{\alpha\beta f}^H \right. \\\nonumber
&&+ \left.{\bf {A}}_q^T \left( ({\bf J}_{pkf} {\bf C}_{pqkf} {\bf J}_{qkf}^H\right.\right. \\\nonumber
&&- \left.{\bf {A}}_p {\bf {X}}\bmath{\Phi}_{\alpha\beta f} \widetilde{\bf C}_{pqkf} \bmath{\Phi}_{\alpha\beta f}^H {\bf {X}}^H{\bf {A}}_q^T)^H {\bf {A}}_p {\bmath \eta}\right.\\\nonumber
&& \left.\left.- {\bf {A}}_q({\bf {X}} \bmath{\Phi}_{\alpha\beta f} \widetilde{\bf C}_{pqkf} \bmath{\Phi}_{\alpha\beta f}^H {\bmath \eta}^H \right.\right.\\\nonumber
&&+\left.\left. {\bmath \eta}\bmath{\Phi}_{\alpha\beta f} \widetilde{\bf C}_{pqkf} \bmath{\Phi}_{\alpha\beta f}^H {\bf {X}}^H)^H {\bf {A}}_p^T{\bf {A}}_p{\bf {X}}\right) \bmath{\Phi}_{\alpha\beta f} \widetilde{\bf C}_{pqkf} \bmath{\Phi}_{\alpha\beta f}^H \right) \\\nonumber.
\eeqn
The only difference is in the gradient with respect to ${\bf Z}$, which is
\beq
\mathrm{grad(L,{\bf Z})} = \sum_j \frac{1}{2}\left(-{\bf Y}_j+\rho\left(-{\bf X}_j + {\bf Z}\right)\right) + \lambda {\bf Z} +\mu \partial \| {\bf Z} \|_1
\eeq
where $\partial \| {\bf Z} \|_1$ is the subgradient of $\| {\bf Z} \|_1$. 
\newpage
\bibliographystyle{IEEEbib}
\bibliography{references}

\end{document}